\pdfoutput=1

\documentclass[sigconf]{acmart}

\renewcommand\footnotetextcopyrightpermission[1]{} 
\usepackage{booktabs} 
\usepackage{amsfonts}
\usepackage{amssymb}
\usepackage{xspace}
\usepackage{epstopdf}
\usepackage{enumitem}
\usepackage{algorithm}
\usepackage{algorithmic}
\usepackage{multirow}
\usepackage{balance}
\usepackage{booktabs}
\usepackage{array}
\usepackage{wrapfig}
\usepackage{pifont}
\usepackage{pbox}
\usepackage{geometry}
\usepackage{color}

\newcommand*\rot{\rotatebox{90}}

\begin{document}

\def\x{{\mathbf x}}
\def\L{{\cal L}}
\def\eg{\textit{e.g.}}
\def\ie{\textit{i.e.}}
\def\Eg{\textit{E.g.}}
\def\etal{\textit{et al.}}
\def\etc{\textit{etc}}

\title{Gender and Emotion Recognition with Implicit User Signals}
\author{Maneesh  Bilalpur}
\affiliation{%
  \institution{International Institute of Information Technology}
  \city{Hyderabad} 
  \state{India} 
}
\email{maneesh.bilalpur@research.iiit.ac.in}

\author{Seyed Mostafa Kia}
\affiliation{%
 \institution{Donders Institute, Radboud University}
  \city{Nijmegen} 
  \state{Netherlands} 
}
\email{s.kia@donders.ru.nl}

\author{Manisha Chawla}
\affiliation{%
  \institution{Centre for Cognitive Science, Indian Institute of Technology}
  \city{Gandhinagar} 
  \country{India}}
\email{manisha.chawla@iitgn.ac.in}

\author{Tat-Seng Chua} 
\affiliation{%
 \institution{School of Computing, National University of Singapore}
 \country{Singapore}}
\email{chuats@comp.nus.edu.sg}

\author{Ramanathan Subramanian}
\affiliation{%
  \institution{University of Glasgow \& Advanced Digital Sciences Center}
  \country{Singapore}
}
\email{ramanathan.subramanian@glasgow.ac.uk}

%
%
%
\renewcommand{\shortauthors}{M. Bilalpur et al.}

\begin{abstract}
We examine the utility of \textbf{\textit{implicit user behavioral signals}} captured using {low-cost}, off-the-shelf devices for anonymous gender and emotion recognition. A user study designed to examine male and female sensitivity to facial emotions confirms that females recognize (especially negative) emotions \textit{quicker} and \textit{more accurately} than men, mirroring prior findings. Implicit viewer responses in the form of \textbf{\textit{EEG brain signals}} and \textbf{\textit{eye movements}} are then examined for existence of (a) emotion and gender-specific patterns from \textit{event-related potentials} (ERPs) and \textit{fixation distributions} and (b) emotion and gender discriminability. Experiments  reveal that (i) Gender and emotion-specific differences are observable from ERPs, (ii) multiple similarities exist between \textit{explicit} responses gathered from users and their \textit{implicit} behavioral signals, and (iii) Significantly above-chance ($\approx$70\%) gender recognition is achievable on comparing emotion-specific EEG responses-- gender differences are encoded best for \textit{anger} and \textit{disgust}. Also, fairly modest \textit{valence} (positive vs negative emotion) recognition is achieved with EEG and eye-based features.   

\end{abstract}

%
%




\maketitle
\section{Introduction}
\textbf{Gender HCI}~\cite{Susan2006} and \textbf{Affective HCI}~\cite{Picard1997} have evolved as critical HCI sub-fields due to widespread acknowledgment of the fact that computers need to \textit{appreciate} and \textit{adapt to} the user's gender and emotional state. The ability to identify user demographics including gender and emotion can benefit interactive and gaming systems in terms of a) visual and interface design~\cite{PassigL01,Czerwinski2002}, (b) game and product recommendation (via ads)~\cite{Zhang16,Homer2012}, and (c) provision of appropriate motivation and feedback for optimizing user experience~\cite{Schwark2013}. Contemporary gender recognition (GR) and emotion recognition (ER) systems primarily work with facial~\cite{Ng2012,joho2011looking} or speech~\cite{Li2013,lee2005toward} cues; however, face and speech are \textit{biometrics} that encode an individual's identity, and pose grave privacy concerns as they can be recorded without the user's knowledge~\cite{privacy}. 

This work examines GR and ER from \textit{\textbf{implicit user behavioral signals}}, in the form of \textit{EEG brain signals} and \textit{eye movements}. The conveyance of implicit behavioral signals is hidden from the outside world, and they cannot be recorded without express user cooperation making them privacy compliant~\cite{Campisi2014}. Also, behavioral signals such as EEG and eye movements are primarily \textit{anonymous} as little is known regarding their uniqueness to a person's identity~\cite{Yang2012}.  

\begin{figure}[th!]
\includegraphics[scale=0.3]{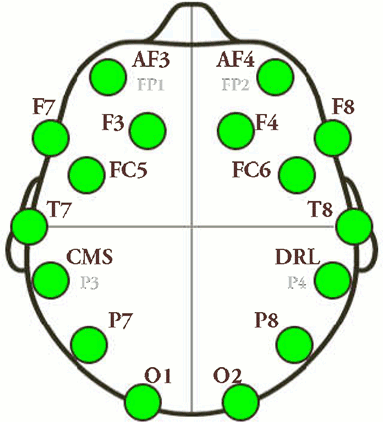}\vspace{-.2cm}
\caption{Emotiv electrode configuration. \label{Emotiv}}\vspace{-.2cm} 
\end{figure}

Specifically, we attempt GR and ER using signals captured by commercial, off-the-shelf devices which are minimally intrusive, affordable and  popularly used in gaming as input or feedback modalities~\cite{Vliet12,LimSW15,etgaming}. The \textit{Emotiv} EEG wireless headset consists of 14 dry (plus two reference) electrodes having a configuration as shown in Fig.~\ref{Emotiv}. While being lightweight, wearable and easy-to-use, neuro-analysis with Emotiv can be challenging due to relatively poor signal quality. Likewise, the \textit{EyeTribe} is a low-cost eye-tracker whose suitability for research purposes has been evaluated and endorsed~\cite{dalmaijer2014low}. We show how relevant gender and emotion-specific information is captured by these low-cost devices via examination of event-related potential (ERP) and fixation distribution patterns, and also through recognition experiments.

To capture gender-based differences, we designed a facial emotion recognition experiment as men and women have been known to respond differently to affective information~\cite{Schwark2013,montagne,halljudith,maneesh2017acii}. Our study performed with 28 viewers (14 male) confirms that women are superior at facial ER, mirroring prior findings. Hypothesizing that enhanced  female emotional sensitivity should reflect via their implicit responses, we examined EEG signals and found that (1) Stronger ERPs are observed for females while processing negative facial emotions, (2) Better ER is achieved with female EEG responses and eye movement features, and (3) Emotion-specific gender differences manifest better than emotion-agnostic differences. 

Overall, our work makes the following research contributions: (a) While prior works have identified gender differences in emotional behavior, this is one of the first works to expressly perform GR and ER from implicit behavioral signals; (b) Apart from recognition experiments, we show that the employed devices capture meaningful information, as typified by gender and emotion-specific event related potentials (ERPs);  (c) The use of minimally intrusive, off-the-shelf and low-cost devices extends the ecological validity of our findings, and the feasibility of employing a similar framework for large-scale interactive user profiling.   

The paper is organized as follows. Section~\ref{RW} reviews related work. Section~\ref{MM} describes the experimental settings and protocol adopted to examine gender differences. Section~\ref{DA} examines \textit{explicit} user responses in terms of response times and recognition rates, while Section~\ref{IC} discusses \textit{implicit} user behavior based on EEG and gazing patterns while also presenting GR and ER experiments. Section~\ref{Con} discusses the salient aspects of this work and concludes the paper.

\section{Related Work}\label{RW}
To position our work with respect to the literature, this section reviews related work on (a) user-centered ER, and (b) gender differences in emotional face processing. 

\subsection{User-centered ER}
\begin{sloppypar}
Emotions evoked by multimedia content have been modeled via \textit{content-centered} as well as \textit{user-centered} approaches. Content-centered approaches typically attempt to find emotional correlates from content-based features~\cite{Hanjalic2005,wang2006affective,Shukla2017acm}, while user-centered methods monitor user behavioral cues (eye movements, EEG signals, \etc) to recognize the evoked emotion. As emotion represents a subjective human feeling, many user-centered approaches have attempted to model emotions by examining \textit{explicit} and \textit{implicit} user behavioral cues. 
Conspicuous facial cues are studied in~\cite{joho2011looking} to detect multimedia highlights, while implicit physiological measurements are employed to model emotions induced by music and movie scenes respectively in~\cite{Koelstra2012} and~\cite{abadi2015decaf}. EEG and eye movements are two popular implicit modalities employed for ER, and many works have predicted affect with a combination of both~\cite{zheng2014multimodal,liu2016multimodal,maneesh2017acii}, or exclusively using either signal~\cite{li2015eeg,zheng2014eeg,Tavakoli15,subramanian2016ascertain,abadi2015decaf,katti2010making,Subramanian2014}. 
\end{sloppypar}

Valence (positive vs negative emotion) recognition using eye-based features is proposed in~\cite{Tavakoli15}. ER with EEG and pupillary responses acquired from five users is discussed in~\cite{zheng2014multimodal}. A deep unsupervised method for ER from raw EEG data is proposed in~\cite{li2015eeg}, and its effectiveness is shown to be comparable to manually extracted features. Differential entropy (DE) features from EEG data are extracted to train an integrated deep belief network plus hidden Markov model for ER in~\cite{zheng2014eeg}. A DAE (differential auto-encoder) that learns shared representations from EEG and eye-based features is proposed for valence recognition in~\cite{liu2016multimodal}. Almost all of these works employ lab-grade eye-trackers and EEG sensors which are bulky and intrusive, and therefore preclude naturalistic user behavior.


\subsection{Gender Differences in Emotion Recognition}
As facial emotions denote important non-verbal communication cues during social interactions, many social psychology studies have examined human recognition of emotional faces. Certain facial features are found to encode emotional cues better than others. When viewing a human face, most attention is drawn to the eyes, nose and mouth regions~\cite{walker-smith,smith}. Authors of~\cite{subramanian2011can} observe that visual attention is localized around the eyes for mildly emotive faces, but the nose and mouth regions also attract substantial eye fixations in highly emotive faces. A recent eye tracking study~\cite{schurgin} notes that distinct eye fixation patterns emerge for different facial emotions. The mouth is most informative for the \textit{joy} and \textit{disgust} emotions, whereas eyes mainly encode information relating to \textit{sadness}, \textit{fear}, \textit{anger} and \textit{shame}. In a similar study~\cite{aviezer}, more fixations are noted on  the upper face half for \textit{anger} as compared to \textit{disgust}, while no differences are observed on lower face half for the two emotions. However, humans can find it difficult to distinguish between some facial emotions that tend to have similar characteristics-- examples are the high overlap rate between the \textit{fear}--\textit{surprise} and \textit{anger}--\textit{disgust} emotion pairs~\cite{smith,etcoff}.

Many studies have also identified gender differences during facial emotion processing. Females are generally found to be better at ER than males, irrespective of age~\cite{sullivan}. Other studies examining the role of facial movements in ER~\cite{bassili,montagne,halljudith} also note that females recognize facial emotions more accurately than males, even when only partial information is available. Some evidence also points to the fact that females are faster at ER than males~\cite{halljessica,rahman}.


Differences in gaze patterns and neural activations have been found between males and females while viewing emotional faces.  Women's tendency to fixate on the eyes positively correlates with their ER capabilities, while men tend to look at the mouth for emotional cues~\cite{sullivan,halljessica}. Likewise, there exists evidence in terms of EEG ERPs that negative facial emotions are processed differently and rapidly by women, and do not necessarily entail selective attention towards emotional cues~\cite{Lithari2010,zotto2015processing}. 

An exhaustive review of GR methodologies is presented in~\cite{wu2015human}, and the authors evaluate GR methods across different communities using metrics like \textit{universality, distinctiveness, permanence} and \textit{collectability}. While crediting bio-signals like EEG and ECG for their accuracy and trustworthiness, this work also highlights the invasiveness of bio-sensors. The sensors used in this work are nevertheless minimally intrusive and non-invasive, thereby ensuring naturalistic user experience, while also recording meaningful emotion and gender-related information. Among the few works to attempt GR from user-centric cues, EEG and speech features are proposed for age and gender recognition in~\cite{nguyen2013age}.

\subsection{Analysis of Related Work}      
A close examination of related work reveals that (1) Many have attempted ER from user-centered cues, both conspicuous and implicit, and a number of works have also tried to isolate gender-specific differences based on gaze patterns and neural activations; nevertheless, very few works have that expressly attempted GR based on implicit user cues. Differently, this work employs implicit signals for GR and ER with a number of classifiers and reliable detection of both gender and valence (AUC score of $>$0.6) is achieved via our proposed methodology; (2) Another salient aspect of our approach is the use of low cost, off-the-shelf and easy-to-use sensors for acquiring implicit signals, which suffer from poor signal quality while enabling natural user behavior. We show how these sensors nevertheless capture meaningful information via the analysis of ERPs and fixation distribution patterns; (3) In contrast to most prior works have either analyzed (a) explicit user responses in terms of reaction times and recognition rates, or (b) implicit behavioral patterns to discover gender differences in ER, or (c) employed EEG and eye-based features for GR and ER without validation of the corresponding signals, our work touches upon all of these aspects. We show how explicit behavioral gender differences noted for negative emotions also reflect via EEG ERPs and fixation distributions, and GR performance. The multiple similarities among explicit and implicit behavioral patterns, along with the use of low-cost sensors validate the findings in this work and extend its ecological validity. The next section describes the experimental set up and protocol employed in our study.      
\section{Materials and Methods}\label{MM}

\subsection{Experimental Overview} We investigated gender differences during visual emotional face processing via explicit recognition rates and implicit EEG plus eye movement patterns. While emotion elicitation can and has been achieved via the presentation of emotional faces to viewers (\eg,~IAPS pictures~\cite{IAPS}), our objective was to study \textit{facial emotion recognition} rather than \textit{affect elicitation}. Specifically, we wanted to validate prior findings regarding the enhanced sensitivity of females to facial emotions. Experimental details are as follows:

\subsubsection{Participants:} 28 subjects from different nationalities (14 male, age $26.1\pm7.3$ and 14 female, age $25.5\pm6$) took part in our study. All of them were university engineering graduate and undergraduate students, and therefore likely possessed similar levels of intellect and social intelligence. All subjects had normal or corrected vision and provided informed consent for participation.

\subsubsection{Stimuli:} 
\begin{figure}[t]
    \centering
   \includegraphics[width=\columnwidth]{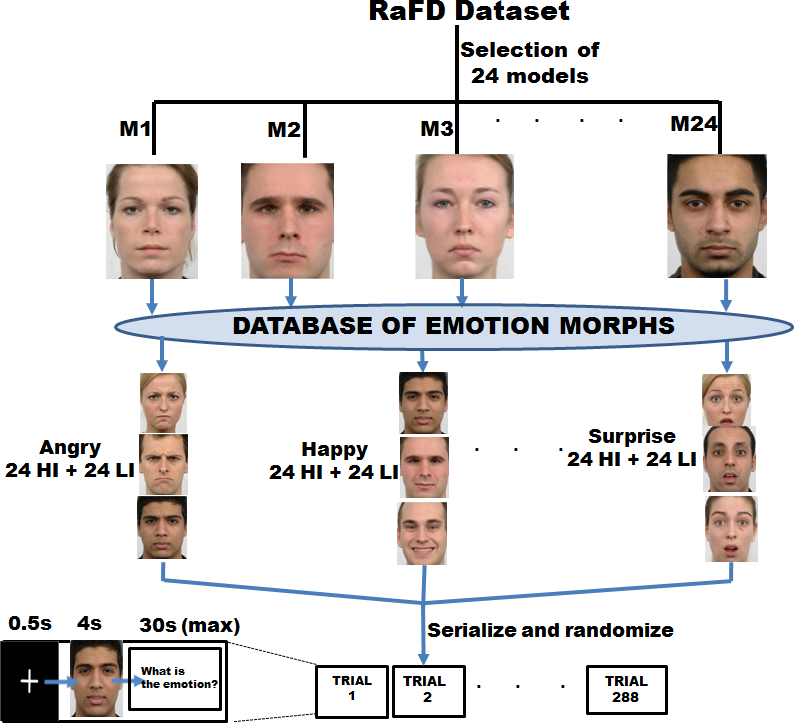}
    \caption{\label{proto} \textbf{Experimental Protocol:} Timeline for each trial is shown on the bottom-left inset. Best-viewed under zoom.}\vspace{-.2cm}
\end{figure}

We used the facial emotions of 24 models (12 male, 12 female) from the Radboud Faces Database (RaFD)~\cite{Rafd}. RaFD contains facial emotions of 49 models which are rated for clarity, genuineness and intensity. 24 models were chosen such that their Ekman facial emotions\footnote{Anger (A), Disgust (D), Fear (F), Happy (H), Sad (Sa) and Surprise (Su).} were roughly matched with respect to the above ratings. We also \textit{morphed} the emotive faces from \textit{neutral} (0\% intensity) to \textit{maximum} (100\% intensity) to generate intermediate morphs in steps of 5\%. Since viewers had difficulty recognizing very low intensity morphs, we only used 25-100\% morphs in our experiments. To examine the impact of morph intensity on ER performance, we divided the morphed emotive faces into \textit{low-intensity} (LI, 25--50\%) and \textit{high-intensity} (HI, 55--100\%) morphs (see Fig.~\ref{proto}). This resulted in a dataset comprising 864 LI and 1440 HI emotional faces. All original and morphed faces were 361$\times$451 pixels in size, translating to a visual angle of 9.1$^\circ$ and 11.4$^\circ$ about $x$ and $y$ at 60 cm screen distance. 

\subsubsection{Protocol:} The experimental protocol was developed using Matlab \textsl{Psychtoolbox}~\cite{Psychtoolbox}, and is outlined in Fig.~\ref{proto}. Corresponding to each model and emotion, we chose one HI and LI morph at random. This generated a total of 288 face images (2 morphs/emotion $\times$ 6 emotions $\times$ 24 models) for presentation to each viewer, and these images were displayed in random order. In each trial (left bottom inset in Fig.~\ref{proto}), an emotive face was displayed for 4s preceded by a fixation cross for 500 ms. Upon image presentation, the viewer had a time-limit of 30s to make \textit{one} out of \textit{seven} choices concerning the facial emotion (6 emotions plus neutral) via a radio button. Neutral faces were only used for morphing and were not presented during the experiment. Viewer EEG responses were recorded via the 14-channel \textit{Emotiv epoc} device, while their eye-movements were captured via the \textit{Eyetribe} eye-tracker. To minimize calibration errors, the experiment was split into 4 parts comprising 72 trials each, and took 90 minutes to complete. 

\section{User response analysis}\label{DA}


We now examine user behavioral data and compare male and female performance in the facial emotion recognition task based on (i) response times (RTs) and (ii) recognition rates (RRs). 



\begin{table*}[ht!]
\caption{ANOVA summary for behavioral measures. df and Sig respectively denote degrees of freedom and significance level.}\label{tab:beh_stat}\vspace{-.2cm}
\centering
\resizebox{0.75\textwidth}{!}{
\begin{tabular}{|c|c|c|c|c|c|c|c|c|}
\toprule	
\textbf{Response}	& \textbf{Predictor} & \multicolumn{1}{|p{0.25\columnwidth}|}{\centering{\textbf{Morph Intensity}}} & \textbf{Emotion} & \textbf{Gender} & \textbf{MI*Emotion}  &\textbf{ MI*Gender} & \textbf{Error} & \textbf{Total} \\ \toprule \bottomrule
	&\textbf{df} & 1 & 5 & 1 & 5 & 5 & 32 & 160 \\ 
	\multirow{ 2}{*}{\textbf{RT}} &\textbf{F} &  5.179 & 7.699 & 3.411 & 2.901 & 3.327 & & \\
	&\textbf{Sig} & $p<0.05$ & $p<0.001$ & $p = 0.074$ & $p<0.05$ & $p<0.05$ & & \\ 
	\multirow{ 2}{*}{\textbf{RR}} &\textbf{F} & 531.1 & 59.83  & 7.37  &  & 7.359 &  & \\
	&\textbf{Sig} & $p<0.001$ & $p<0.001$ & $p <0.05$ &  & $p<0.001$ &  & \\ \hline
	\end{tabular}}
\end{table*}

\subsection{Response time (RT)}\label{RT}
\begin{figure}[t]
\includegraphics[width=1.0\linewidth]{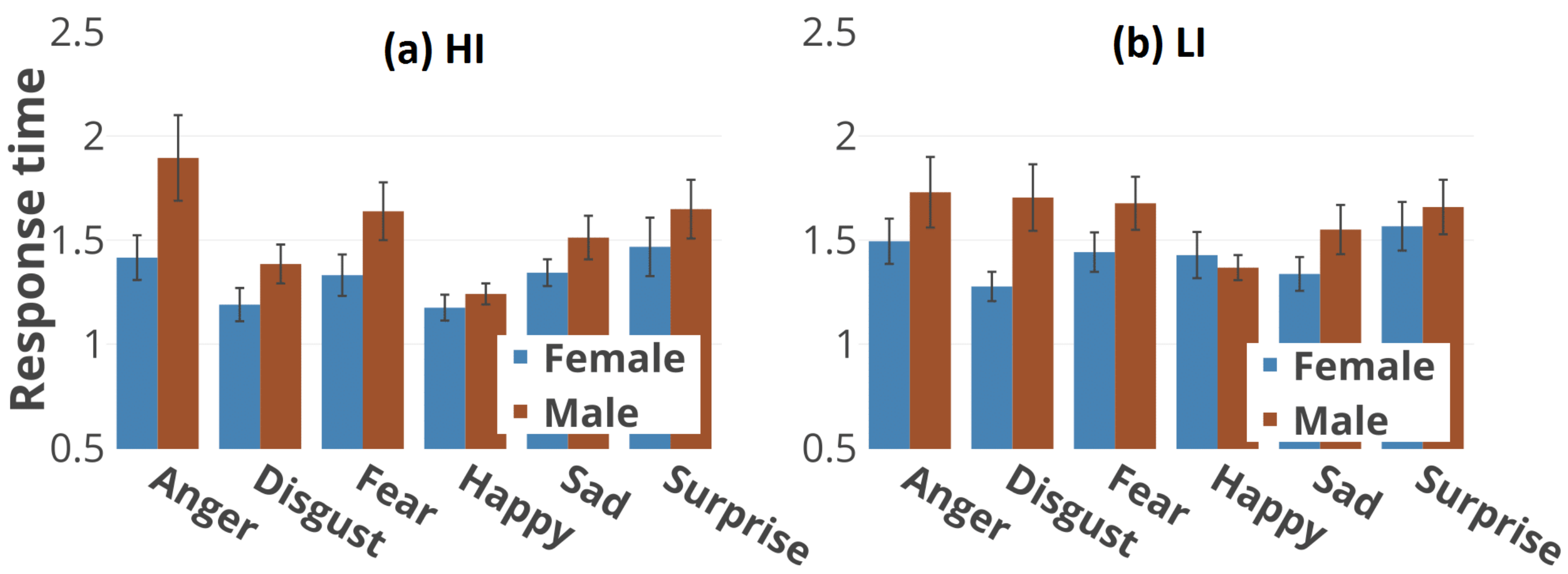}
\caption{\label{RT_HR} Emotion-wise mean RTs of males and females considering HI and LI  morphs. Error bars denote unit standard error (best-viewed in color).}
\vspace{-.1cm}
\end{figure}

The overall viewer RT was 1.5$\pm$0.07 s, implying that users judged facial emotions fairly instantaneously. Table~\ref{tab:beh_stat} summarizes results from a 3-way ANOVA on RTs, with \textit{\textbf{morph intensity}} (MI) and \textit{facial \textbf{emotion}} as within-subject factors, and \textit{\textbf{gender}} as the between-subject factor. ANOVA revealed the main effect of MI with generally faster RTs noted for HI morphs (see Fig.~\ref{RT_HR}). The presented facial motion also impacted RTs, and pairwise comparisons revealed that RT for \textit{happy} ($1.3\pm0.9$ s) was significantly faster than for \textit{anger}, \textit{fear}, \textit{sad} and \textit{surprise}. ANOVA also revealed a marginal gender effect with females ($1.4\pm0.1$ s) generally responding faster than males ($1.6\pm0.1$ s). Significant interaction effects were also noted between (a) morph intensity and emotion, and (b) gender and emotion. 
As evident from Fig.~\ref{RT_HR}, females were significantly faster ($F(3,96)= 3.33, p < 0.05$) at judging negative (\textit{anger}, \textit{disgust}, \textit{fear} and \textit{sad}) emotions.

\subsection{Recognition rates (RRs)}\label{RR}
\begin{figure}[t]
\centering
\includegraphics[width=1.0\linewidth]{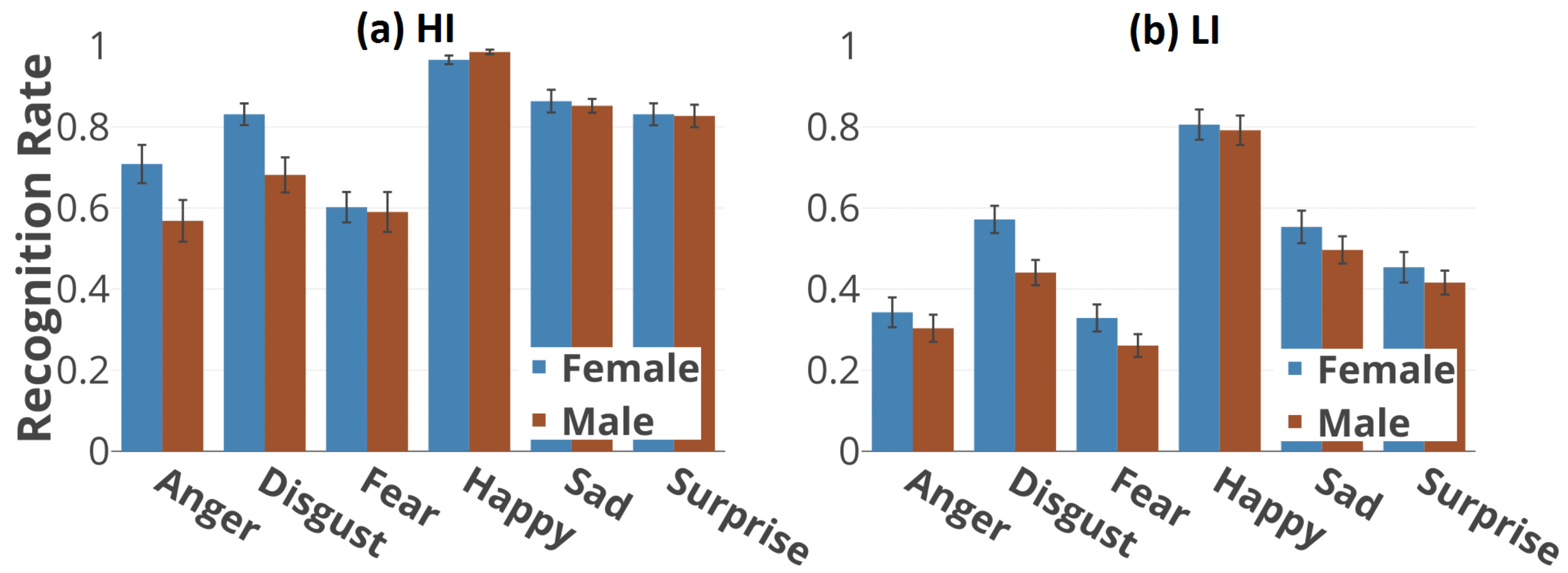}
\caption{\label{RRnfix} Male and female RRs for HI morphs, and LI morphs.} \vspace{-.1cm}
\end{figure}

A 64\% RR was achieved overall, suggesting that facial emotions were recognized well above chance level but well below ceiling. ANOVA summary in Table~\ref{tab:beh_stat} reveals the main effect of morph intensity, emotion and gender. Females were more accurate than males at recognizing negative emotions, even from LI morphs (Fig.~\ref{RRnfix}(a,b)). HI morphs were recognized more accurately (mean RR = 78.7) than LI morphs (mean RR = 49.1). Also, \textit{happy} faces were recognized most accurately, and \textit{fearful} faces least accurately.  We also found a significant interaction effect between morph intensity and emotion. Overall, (a) HI emotion morphs were recognized way more accurately than LI morphs, (b) females recognized negative emotions better than males and (c) \textit{Happy} was the easiest emotion to recognize, with \textit{fear} being the most difficult. 

%



\section{Analyzing Implicit Cues}\label{IC}

Having noted that females are quicker and better at recognizing (especially negative) emotions, we hypothesized that these behavioral differences should also manifest via implicit cues in the form of brain responses and eye movements during emotional face processing. This section describes the features extracted and methods employed to examine these cues.

\subsection{EEG analysis}
We extracted epochs for each \textit{trial} (4.5s of stimulus-plus-fixation cross viewing time @ 128 Hz sampling rate, see Fig.~\ref{proto}), and the 64 pre-stimulus fixation samples were used to remove DC offset. This was followed by (a) EEG band-limiting to within 0.1--45 Hz, (b) removal of noisy epochs via visual inspection, and (c) independent component analysis (ICA) to remove  artifacts relating to eye-blinks, and eye and muscle movements. Muscle movement artifacts in EEG are mainly concentrated between 40--100 Hz~\cite{Barua3562,Suresh13,Gasser20052044}. While most artifacts were removed upon EEG band-limiting, the remaining were removed manually via inspection of ICA components. Finally, 7168 (14 electrodes $\times$ 128 Hz $\times$ 4s) dimensional feature matrices were input to the classifier following dimensionality reduction via principal component analysis (PCA) to retain 90\% input variance.

\begin{figure*}[t]
\includegraphics[width=0.48\linewidth]{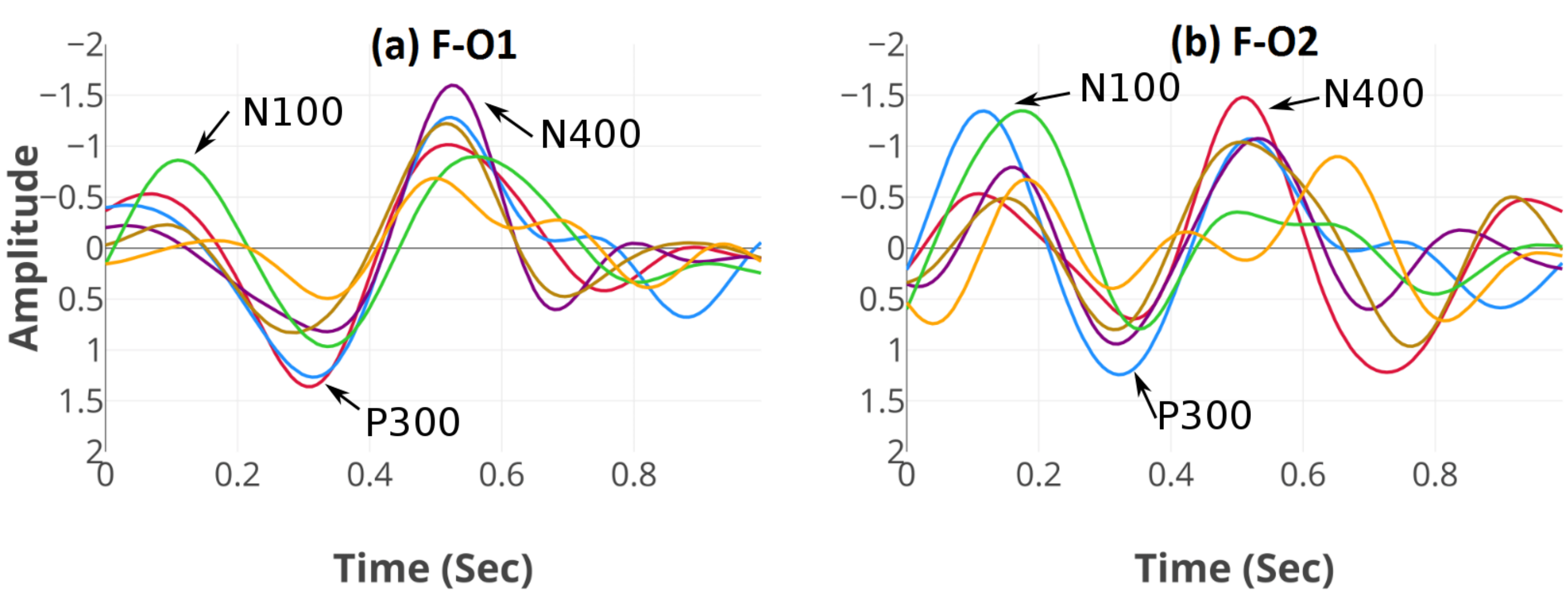}
\includegraphics[width=0.48\linewidth]{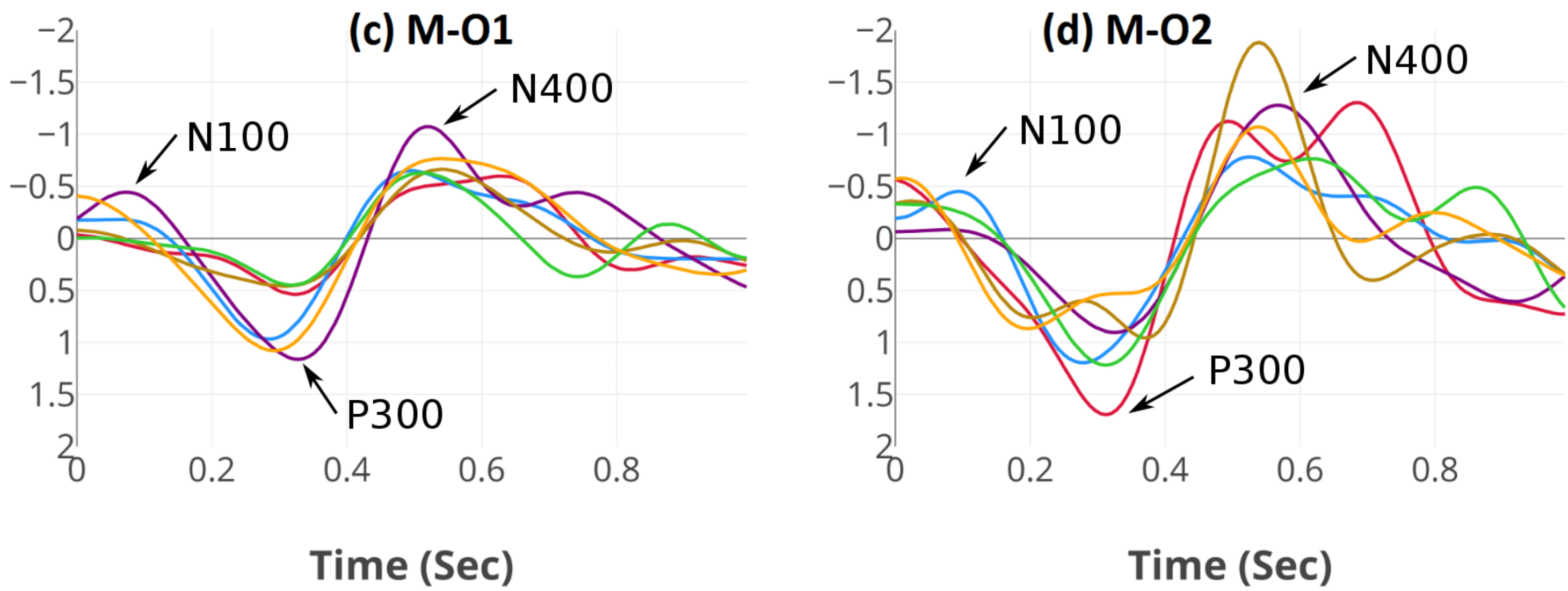}
\vspace{-2cm}
\includegraphics[width=0.48\linewidth]{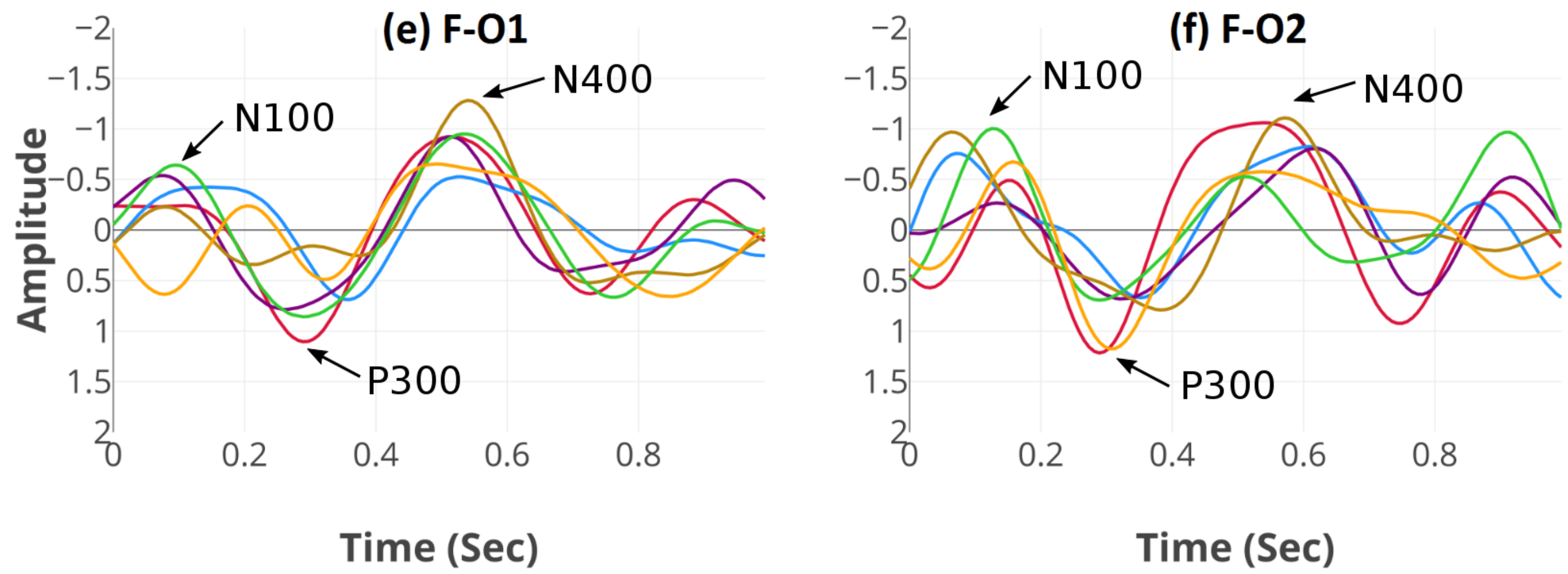}
\includegraphics[width=0.48\linewidth]{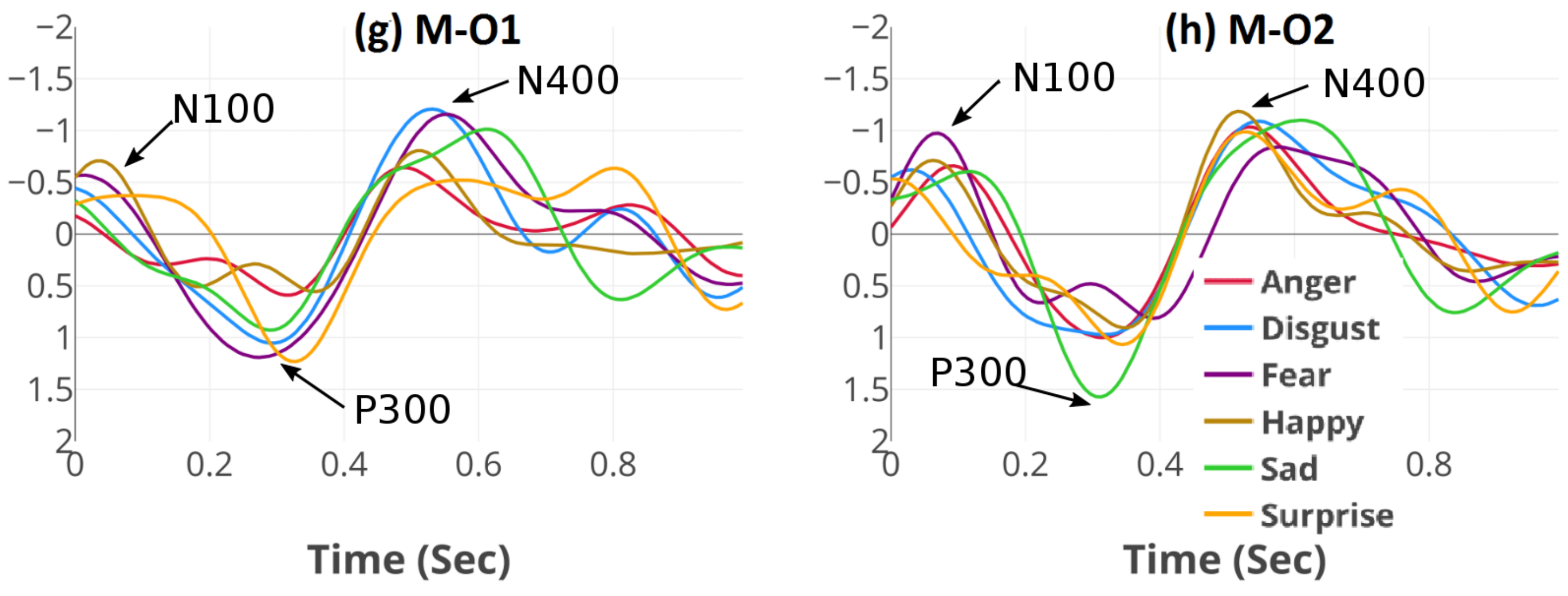}
\vspace{2cm}
\caption{\label{ERP} ERPs for HI  morphs (a--d) and LI morphs (e--h): (from left to right) O1 and O2 ERPs for females and males.$y$-axis shows ERP in microvolts, refer (h) for legend. ERP data is plotted upside down (view in color and under zoom).} \vspace{-.2cm}
\end{figure*}

\subsubsection{ERP analysis}\label{ERPsec}
Prior works have observed event-related potential\footnote{ERP denotes \textit{average} EEG response over multiple trials. P300 (positive) and visual N100, N400 (negative) are examples of ERP components which respectively peak around 300, 100 and 400 ms post stimulus onset.} (ERP) based gender differences from EEG data recorded with lab-grade sensors~\cite{Lithari2010,Muhl14,zotto2015processing}. Specifically,~\cite{Lithari2010} notes enhanced negative ERPs for females in response to negative emotional stimuli. However, capturing ERPs with commercial devices is challenging due to their low signal-to-noise ratio. Fig.~\ref{ERP} presents the P300, visual N100 and N400 ERP components in the occipital O1 and O2 electrodes (see Fig.~\ref{Emotiv}) corresponding to the HI and LI faces. Note that the occipital lobe represents the \textit{visual processing center} in the brain, as it contains the primary visual cortex. 

Comparing O1/O2 male and female ERPs for positive (H,Su) and negative (A,D,F,Sa) emotions, no significant difference can be observed between male positive and negative ERP peaks for HI or LI faces (see columns 3,4). However, we observe stronger N100 and P300 peaks in the negative female ERPs for both HI and LI faces (columns 1,2). Also, a stronger female N400 peak can be noted for HI faces consistent with prior findings~\cite{Lithari2010}. Lower male N100 and P300 latencies can be observed for positive HI emotions, with the pattern being more obvious at O2. Also, lower male N400 latencies can be noted at O2. The positive vs negative ERP difference for females is narrower for LI faces, which may be attributed to the greater difficulty in identifying LI emotions. This is further corroborated by the fact that LI faces generally produce weaker ERPs at O1 and O2 than HI faces. Overall, the observed ERPs confirm that gender differences in emotional face processing can be effectively captured by the \textit{Emotiv} EEG device. 

\subsection{Eye-tracking analysis}
Gender differences in gaze patterns during emotional face processing have been observed by prior works~\cite{schurgin,sullivan} using high-end eye trackers. In this work, we used the low-cost \textit{Eyetribe} device with 30Hz sampling rate to synchronously record eye movements along with the EEG signals, and compare male and female eye movement patterns. To compute \textit{fixations} from raw gaze positions estimated by the tracker, we used the EyeMMV toolbox~\cite{krassanakis2014eyemmv} and considered the transition from one fixation to another as a \textit{saccade}. Upon extracting fixations and saccades, we adopted the features employed for valence recognition in~\cite{Tavakoli15}, namely, saccade orientation, top-ten salient coordinates, saliency map and histograms of a) saccade slope, b) saccade length, c) saccade velocity, d) fixation duration, e) fixation count and f) saliency to compute an 825-dimensional feature vector for our analyses.

\subsubsection{Fixation analysis}
\begin{figure}[t]
\centering
\includegraphics[width=1.0\linewidth]{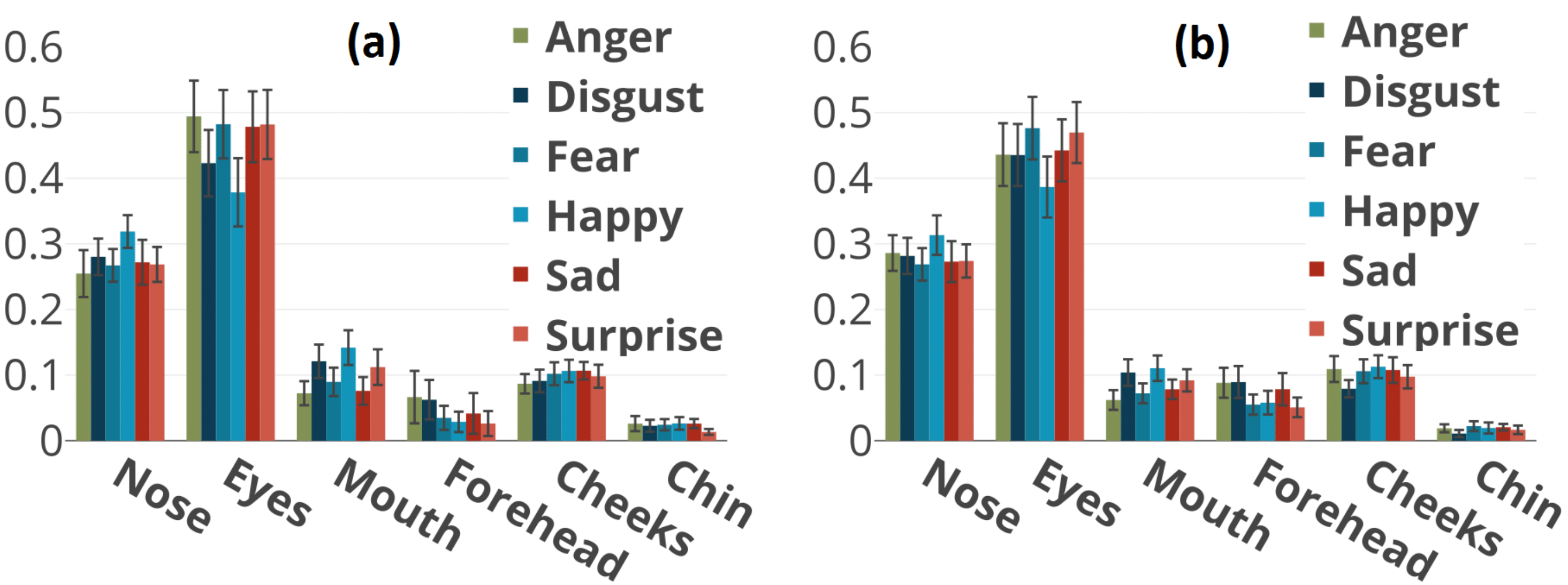}
\vspace{-0.8cm}
\caption{Fixation duration distributions: Region-wise fixation distribution across different emotions for (a) females and (b) males. Error bars denote unit standard error. \label{Fix_dist}}\vspace{-.2cm} 
\end{figure}

To validate observations on the fixating patterns of males and females, we computed the fixation duration distribution over six facial regions, namely, eyes, nose, mouth, cheeks , forehead and chin. Fig.~\ref{Fix_dist} presents the males and female fixation duration distribution over these facial regions for the various emotions considering (HI$+$LI) faces. For both men and women, the time spent examining the eyes, nose and mouth account for over 80\% of the total fixation duration, with eyes ($\approx$45\%) and nose ($\approx$30\%) attracting maximum attention as observed in \cite{schurgin}. Focusing on the emotion-wise distributions, a high proportion of eye fixations is noted for all emotions except \textit{happy}, while relatively higher visual attention is noted on the nose and mouth for this emotion consistent with the findings in~\cite{schurgin} implying that these regions encode key emotional information concerning happiness. Overall, females were found to examine the eyes and mouth more for emotional cues, with males fixating more on the nose even though none of the differences were significant. The next section presents ER and GR results achieved with EEG and eye-based features. 

\subsection{Experiments and Results}
\subsubsection{Experimental settings} We separately examined EEG and eye-based responses to HI and LI morphs for emotion and gender recognition. Specifically, we evaluated ER and GR performance upon feeding (a) only EEG, (b) only eye and (c) concatenated EEG$+$ eye features (termed \textit{early fusion} or EF) to classifiers. Additionally, we probabilistically fused the EEG and eye-based classifier outputs (termed \textit{late fusion} or LF) via the $W_{est}$ procedure outlined in~\cite{koelstra2012fusion}, and briefly described as follows. From the EEG and eye-based classifier outputs, the test sample is labeled based on the larger $P_i = \sum_{i=1}^2 \alpha_i^*t_ip_i$, where $i$ indexes the two modalities, $p_i$'s denote posterior individual classifier probabilities for the `high' and `low' classes (we only consider binary GR and ER) and $\{\alpha_i^*\}$ are the optimal weights maximizing test performance, as determined via a 2D grid search. If $F_i$ denotes training performance for modality $i$, then $t_i = \alpha_i F_i/\sum_{i=1}^2 \alpha_i F_i$ for given $\alpha_i$. 

\paragraph*{Classifiers and training data} For ER and GR, we considered the Naive-Bayes (NB), linear SVM (LSVM) and radial-basis SVM (RSVM) classifiers. NB is a linear generative classifier that assumes class-conditional feature independence to estimate the posterior $p(C/X)$, where $C$ and $X$ denote class-label and feature vector for the test sample. LSVM and RSVM respectively denote the linear and radial basis kernel versions of the popular SVM classifier. For both ER and GR, we only fed (EEG and eye) data corresponding to \textit{correct} trials where users correctly recognized the presented facial emotion. 

\paragraph*{Performance measure} For benchmarking ER/GR performance, we considered the area under curve ({\textbf{AUC}}) performance measure. The AUC represents area under the ROC curve plotting true and false positive rates, and a random classifier will correspond to an AUC score of 0.5. Also, AUC is an appropriate metric while evaluating classifier performance on data with unbalanced positive and negative class sizes. As we are attempting recognition with few training data, we report ER/GR results over five repetitions of 10-fold cross validation (CV) (\ie, total of 50 runs). CV is typically used to overcome the \textit{overfitting} problem on small datasets, and the optimal SVM parameters were determined from the range $[10^{-4},10^{4}]$ via an inner 10-fold CV on the training set.  

\subsubsection{Emotion Recognition} \label{ER}
As explicit behavioral responses (Sec.\ref{DA}) and ERP-based (Sec.\ref{ERPsec}) analyses suggest that (a) females are especially sensitive to negative facial emotions, and (b) there are cognitive differences in the processing of positive and negative emotions for females and males, we modeled ER as a binary classification problem, with an objective to distinguish between positive and negative  emotional classes.

\begin{figure}[t]
\includegraphics[width=1.0\linewidth]{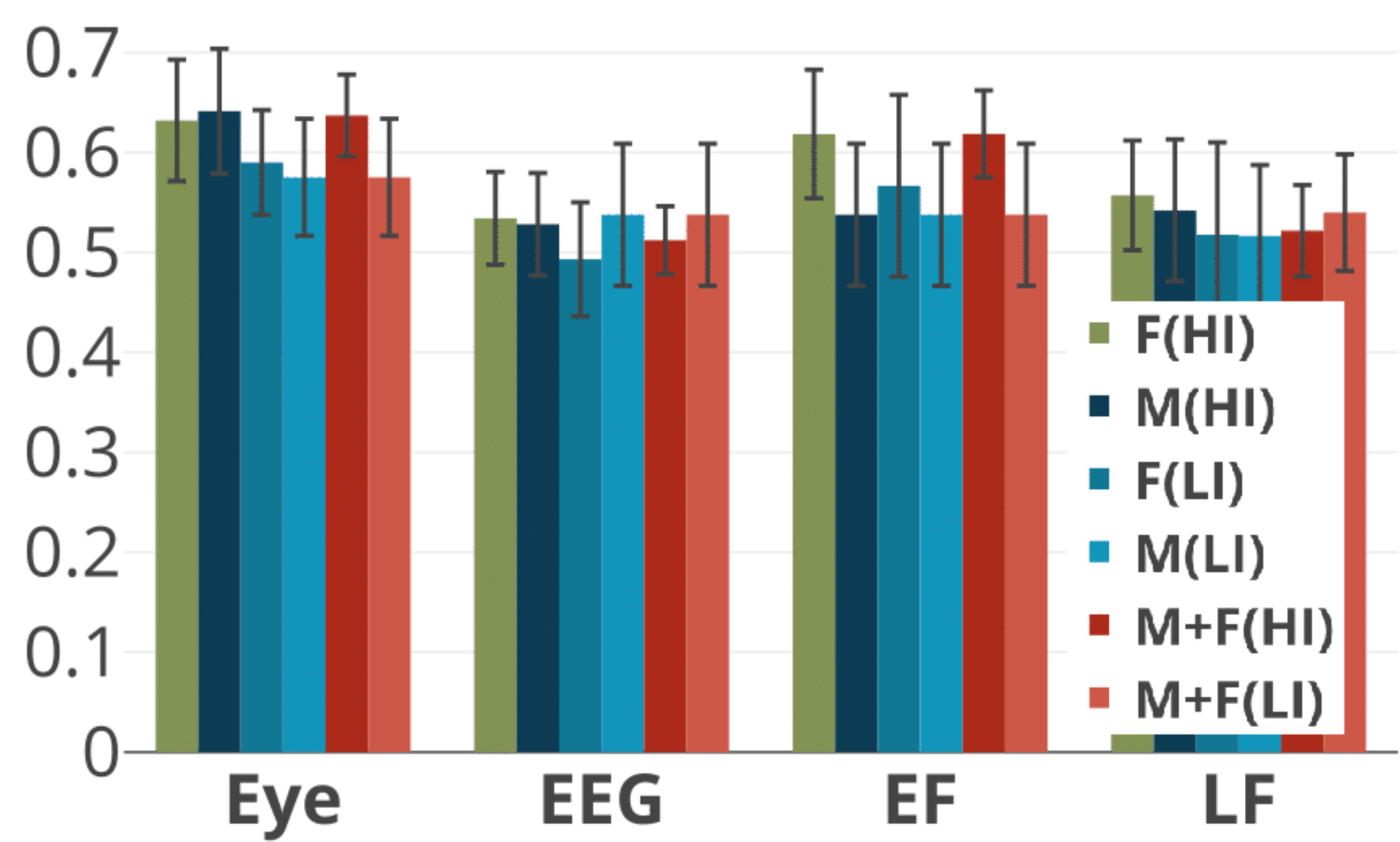}
\vspace{-0.8cm}
\caption{ER Results: valence recognition results with HI and LI morphs.
\label{Rec_Emotiv}}\vspace{-.5cm} 
\end{figure}

Fig.~\ref{Rec_Emotiv} presents valence recognition (VR) results obtained with male, female and male$+$female unimodal and multimodal features. Expectedly, better user-centered VR is achieved overall on examining responses to HI emotions (minimum AUC = 0.51 for HI vs 0.49 for LI morphs). Also, best VR performance is achieved with female data (mean AUC = 0.58 with HI morphs) as compared to male (mean AUC = 0.56, HI morphs) and male$+$female data (mean AUC = 0.543 for HI morphs) considering both unimodal and multimodal features. Interestingly, eye-based features (peak AUC = 0.64 with male data) are more discriminative for VR than EEG features (peak AUC = 0.53 with female data). Evidently, fusing the eye and EEG information is not beneficial even though a simple concatenation of the EEG and eye features is found to be superior to 
probabilistic fusion of unimodal results.

The obtained results are competitive against (and in some cases better than) prior approaches. Eye based features are found to achieve 52.5\% valence recognition accuracy in~\cite{Tavakoli15}, where emotions are induced by presenting diverse types of emotional scenes to viewers while only emotional faces employed in this work. Also, prior neural studies~\cite{abadi2015decaf,Muhl14} which employ music and movie content for emotion elicitation achieve only around 60\% VR with lab-grade sensors. Our VR results are creditable especially considering that low-cost EEG and eye tracking devices are used in this work and that our objective was not to induce emotions in viewers \textit{per se}-- the discriminability instead arises from the manner in which facial emotions are \textit{perceived} by viewers. Superior performance of eye-based features may be attributed to the fact that in addition to the difference in fixation duration observable for \textit{happy} (Fig.~\ref{Fix_dist}), the saccade and saliency based statistics are able to capture discrepancies in gazing patterns for positive and negative facial emotions~\cite{Tavakoli15}; in contrast, EEG data are vectorized and analyzed \textit{as is} without any explicit feature extraction.  
  
%

\subsubsection{Gender Recognition}
Table~\ref{GenResults} presents eye and EEG-based GR results obtained with HI and LI emotions. Specifically, GR AUC scores with various modalities and classifiers\footnote{with which optimal GR results are achieved for a given condition} corresponding to \textit{all} and \textit{emotion-specific} trials are tabulated. Clearly better than chance GR is achievable with both EEG and eye-based features. An immediate observation from the tabulated results is that unlike for ER, very similar GR performance is noted for both HI and LI emotions, implying that gender specific differences are also effectively encoded in implicit viewer responses to mildly emotive faces. This is consistent with the explicit user behavioral results in Fig.\ref{RRnfix}, where women are found to outperform men at recognizing LI facial emotions. 


\begin{table}[bh!]
\centering
\resizebox{0.475\textwidth}{!}{
\begin{tabular}{l|l|c|cc}
\hline
\multicolumn{3}{c|}{\textbf{AUC}} & \textbf{HI} & \textbf{LI} \\
\hline
\multicolumn{2}{c|}{\multirow{4}{*}{\textbf{All}}} &
EEG(NB) & \textbf{0.714 $\pm$ 0.002} & \textbf{0.600 $\pm$ 0.005} \\
\multicolumn{2}{c|}{} & EYE(NB) & 0.493 $\pm$ 0.013 & 0.481 $\pm$ 0.017 \\
\multicolumn{2}{c|}{} & Early fusion(RSVM) & 0.522 $\pm$ 0.035 & 0.524 $\pm$ 0.022\\
\multicolumn{2}{c|}{} & Late fusion(LSVM) & 0.549 $\pm$  0.022 & 0.523 $\pm$ 0.035\\
\hline
\multirow{24}{*}{\rot{\textbf{Emotion wise}}} & \multirow{6}{*}{\rot{\textbf{EEG (NB)}}} & A &
\textbf{0.708 $\pm$ 0.064} & 0.580 $\pm $0.074 \\
 &  & D & 0.673 $\pm$ 0.055 & \textbf{0.696 $\pm$ 0.062} \\
 &  & F & 0.643 $\pm$ 0.059 & 0.596 $\pm$ 0.089 \\
 &  & H & 0.696 $\pm$ 0.047 & 0.668 $\pm$ 0.046 \\
 &  & Sa & 0.674 $\pm$ 0.048 & 0.634 $\pm$ 0.064 \\
 &  & Su & 0.692 $\pm$ 0.048 & 0.636 $\pm$ 0.071 \\
\cline{2-5} 
 & \multirow{6}{*}{\rot{\textbf{EYE (NB)}}} & 
A & \textbf{0.601 $\pm$ 0.021} & 0.565 $\pm$ 0.031 \\
 &  & D & 0.577 $\pm$ 0.011 & \textbf{0.632 $\pm$ 0.009} \\
 &  & F & 0.595 $\pm$ 0.015 & 0.535 $\pm$ 0.029 \\
 &  & H & 0.560 $\pm$ 0.021 & 0.538 $\pm$ 0.017 \\
 &  & Sa & 0.539 $\pm$ 0.015 & 0.605 $\pm$ 0.030 \\
 &  & Su &0.555 $\pm$ 0.008 & 0.555 $\pm$ 0.018 \\
\cline{2-5} 
 & \multirow{6}{*}{\rot{\textbf{EF (RSVM)}}} & 
A & 0.555 $\pm$ 0.021 & 0.581 $\pm$ 0.037 \\
 & & D & 0.535 $\pm$ 0.024 & \textbf{0.622 $\pm$ 0.025} \\
 & & F & \textbf{0.618 $\pm$ 0.011} & 0.619 $\pm$ 0.041 \\
 & & H & 0.575 $\pm$ 0.017 & 0.597 $\pm$ 0.009 \\
 & & Sa & 0.540 $\pm$ 0.021 & 0.598 $\pm$ 0.024 \\
 & & Su & 0.579 $\pm$ 0.013 & 0.574 $\pm$ 0.014 \\
\cline{2-5} 
 & \multirow{6}{*}{\rot{\textbf{EF (RSVM)}}} &
A & 0.543 $\pm$ 0.062 & 0.571 $\pm$ 0.081 \\
 & & D & 0.542 $\pm$ 0.044 & \textbf{0.597 $\pm$ 0.093} \\
 & & F & 0.519 $\pm$ 0.029 & 0.597 $\pm$ 0.170 \\
 & & H & 0.526 $\pm$ 0.031 & 0.508 $\pm$0.017 \\
 & & Sa & \textbf{0.573 $\pm$ 0.076} & 0.584 $\pm$ 0.102 \\
 & & Su & 0.562 $\pm$ 0.073 & 0.568 $\pm$ 0.125 \\
\hline
\end{tabular}}
\vspace{0.5cm}
\caption{Table showing Gender Recognition results over different modalities and their fusion for HI and LI morphs using NB-NaiveBayes,RSVM-SVM with RBF kernel and LSVM-SVM with linear kernel.}
\label{GenResults}
\end{table}


Focusing on specifics, EEG features considerably outperform eye-based features for GR and higher AUC scores are obtained with \textit{emotion-specific} (exclusively negative emotion) features as compared to \textit{emotion agnostic} features. This finding is consistent with Fig.\ref{RRnfix} and prior works such as~\cite{Lithari2010,zotto2015processing} which have noted significant differences in female neural responses to negative facial emotions. Particularly, the best EEG AUC scores for HI and LI morphs are obtained for the anger and disgust emotions respectively, for which the greatest difference in recognition rates (of 0.14 and 0.13) can be observed from Fig.\ref{RRnfix}. These findings are salient and interesting, as Fig.\ref{RRnfix} and Table~\ref{GenResults} present two different phenomena, namely, gender differences in ER and discriminability of male and female EEG patterns.

Surprisingly, multimodal fusion performs inferior to EEG/eye features. Also, marginally superior GR results are achieved with early fusion, suggesting that perhaps little complementary information is encoded between the EEG and eye-based features. Finally, RSVM is found to produce the best GR performance in most conditions implying that the extracted EEG and eye features are better separable in a higher-dimensional feature space. 

\subsubsection{Spatio-temporal EEG analysis} As EEG represents multi-channel time-series data, we also performed spatial and temporal analyses on the EEG signals to discover (i) the key EEG electrodes that capture gender differences in emotional face processing, and (ii) the critical time window for GR over the 4s of stimulus presentation (or visual emotional processing).

\begin{figure*}[th!]
\centering
\vspace{-0.5cm}
\includegraphics[width=1.0\linewidth]{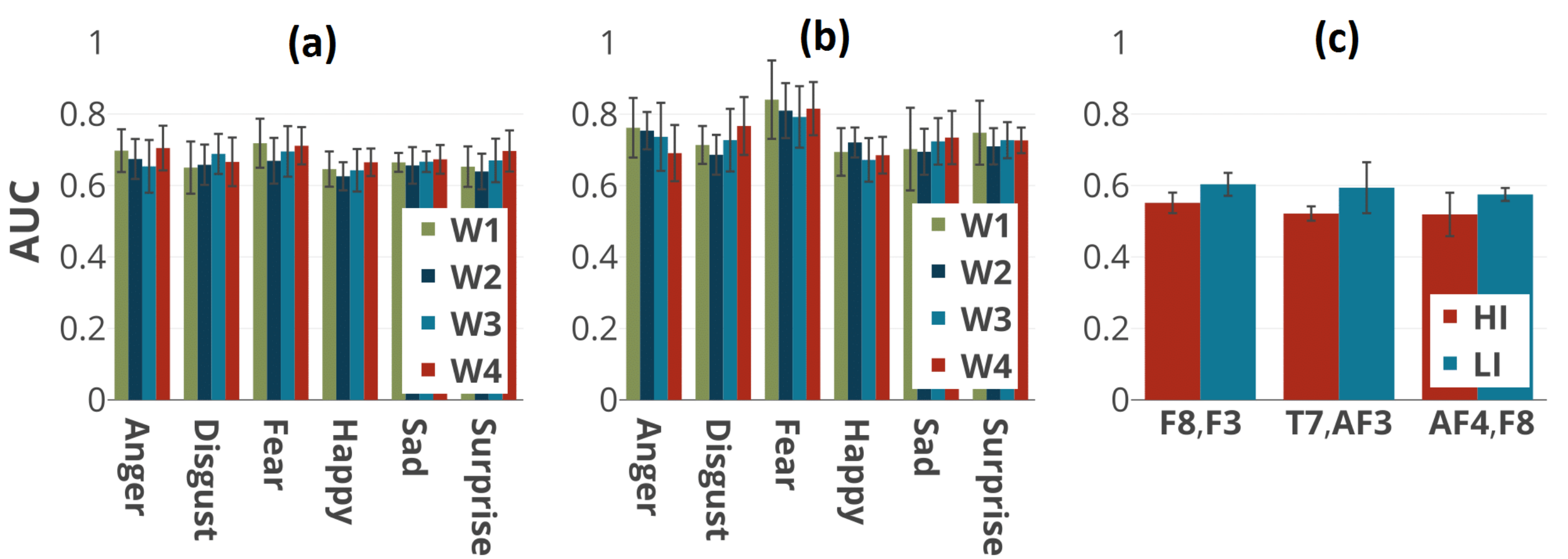}
\vspace{-0.7cm}
\caption{(a,b) Spatio-temporal EEG analyses: GR results for HI  and LI morphs over temporal windows (W1--W4). (c) Top 3 electrodes based on AUC metric for HI and LI morphs. Error bars denote unit standard error.} 
\label{winGenResults}
\vspace{-.1cm}
\end{figure*}

Fig.\ref{winGenResults}(a,b) show the temporal variance in GR performance for HI and LI morphs when the 4s emotional face viewing time is split into four 1s-long windows. These results are computed considering emotion-specific EEG epochs and with information from all 14 electrodes. Overall, no consistent temporal trends concerning gender discriminability can be noted from the plots. However, higher AUC scores over temporal windows  W1--W4 are achieved with LI morphs as compared to HI morphs. Also, slightly higher AUC metrics are noted for negative emotions especially with LI morphs, and the best and worst scores are obtained for the \textit{fear} and \textit{happy} emotions. Results of spatial analyses to identify the top three electrodes that best encode gender differences in emotion processing (EEG data over all 4s is used for this analysis.) are shown in Fig.\ref{winGenResults}(c). Again, superior GR is noted with LI morphs in this scenario, and five of the six electrodes with the best AUC scores correspond to the \textit{frontal} lobe. This result is consistent with the findings in ~\cite{Mclure04}, which notes that gender differences in emotion processing are best encoded in the prefrontal cortex.   

\section{Discussion \& Conclusion}\label{Con}
Of late, there has been considerable interest in the use of technologies such as EEG and eye-tracking to study the cognitive and emotional behavior of users via implicit behavioral cues. Even as potential applications range from affect estimation~\cite{Koelstra2012,abadi2015decaf}, prediction of erroneous actions in collaborative settings~\cite{Vi2014} and evaluating the design of data visualization interfaces~\cite{peck2013using}, many of these studies have been limited to lab settings with bulky and intrusive sensors constraining the manner in which users can express themselves. 

A key impediment towards employing off-the-shelf sensing devices, which enable naturalistic user behavior in interactive applications has been their poor signal quality, and skepticism regarding the nature of `useful' information they actually encode. In this regard, we believe that this work adds on to a small body of works such as~\cite{Vi2014} that have demonstrated the encoding of meaningful information with commercial sensors.    

In particular, our work employs the wireless and wearable \textit{Emotiv} EEG device along with the low-cost \textit{Eyetribe} eye tracker to discover gender differences in cognitive and gazing behavior during a facial emotion recognition task. The validity of our study is endorsed by the multiple correlations observed between \textit{explicit} user responses and their \textit{implicit} behavioral patterns. To begin with, women are found to recognize negative facial emotions quickly and more accurately in our experiment, and correspondingly, higher female ERP amplitudes are noted for negative emotions; peak GR performance with EEG, eye features and their fusion is also noted for negative emotions. The maximum difference in female and male recognition rates for HI and LI morphs are noted for the angry and disgust emotions, and the peak GR AUC scores are also noted for these emotions. 

Overall, reliable GR with EEG (peak AUC of 0.71), eye features (peak AUC of 0.63) and their fusion (peak AUC of 0.62), as well as valence recognition (peak AUC of 0.64 with eye data) is achieved using the proposed framework. The use of minimally intrusive and low-cost devices for studying user behavior makes our findings ecologically valid, and enhances the possibility of large-scale user profiling. 
However, limitations in preprocessing like ICA and visual rejection that require a trained eye exist. An alternate to this could be to adopt machine learning methods like deep learning which is now providing end-to-end system architectures with feature extraction incorporated.

The larger aim of this work is to predict user demographics via implicit and anonymous behavioral cues so as to implement intelligent advertising, recommendation systems and mental health monitoring systems(for disorders like Alexithymia).Some of the applications are described here. A recommendation system could be built by evaluating user engagement in advertisements/movies-- movie reviews can be highly impacted by certain sequences in it, such reviews are often biased and do not provide a holistic feedback of the entire movie. Using our emotion recognition framework, one can identify the sequences where the audience were distracted, for a better rating system and a feedback system for an immersive cinematic experience. Emotional human faces
(especially happy and anger) are ubiquitous on e-commerce websites specifically in categories like
clothing and accessories. These can act as stimuli to obtain EEG response and predict the gender 
to recommend appropriate products. As the window based analysis has shown, the stimulus time could 
be reduced to close to a second which is practically more feasible system so as to speedup the GR
and provide a smooth customer experience. This work represents a proof of concept study, and one can expect more robust and accurate results with larger training data and sophisticated machine learning approaches. Furthermore, computers need to understand and adapt to user emotions for effective human-computer interaction, and gender-specific emotional aspects can be effectively utilized by affective interfaces for manipulating user behavior.

\section*{Acknowledgement}
\noindent This study is partly supported by the Human Centered Cyber-physical Systems research grant from Singapore's Agency for Science, Technology and Research (A*STAR).



\bibliographystyle{ACM-Reference-Format}
\bibliography{sample}

\end{document}